\documentclass[twocolumn,showpacs,floatfix,preprintnumbers,amsmath,amssymb]{revtex4}

\usepackage{graphicx}

\begin{document}

\preprint{APS/123-QED}

\title{Dependance of weak localization cone with the photons brownian Doppler shift}
\author{Max Lesaffre, Michael Atlan}
\author{Michel Gross}\email{gross@lkb.ens.fr}
\affiliation{Laboratoire Kastler Brossel de L'Ecole Normale
Sup\'erieure
 UMR 8552 CNRS, 24 rue Lhomond F-75005 Paris
}

\date{\today}

\begin{abstract}
We report the first observation of the  dependence of the coherent
backscattering (CBS) enhanced cone with the frequency of the
backscattered photon. The experiment is performed on a diffusing
liquid suspension and the Doppler broadening of light is induced
by the brownian motion of the scatterers. Heterodyne detection on
a CCD camera is used to measure the complex field (i.e. the
hologram) of the light that is  backscattered at a given
frequency. The analysis of the holograms yield the frequency and
the propagation direction of the backscattered photons. We observe
that the angular CBS cone becomes more narrow in the tail of the
brownian spectrum. The experimental results are in good agreement
with a simple theoretical model.
\end{abstract}


\maketitle

DOI: 10.1103/PhysRevLett.97.033901

\bigskip

Coherent backscattering (CBS) of light is a photon self
interference effect which leads to an enhanced intensity cone in
the backscattering direction. This effect, which is related to the
solid state physics phenomenon of weak Anderson localization, has
been extensively studied both experimentally and theoretically.
Since the first observation of CBS from colloidal suspension
\cite{albada_85,Wolf_85}, the phenomenon has been observed on many
experimental systems such as  powders
\cite{etemad_86,Kaveh_86,Wiersma_95}, liquid crystals
\cite{Vlasov_88},  photonic crystals \cite{Koenderink_2000}, and
cold atoms gases
\cite{Labeyrie_99,Chanelière_04,Shatokhin_05,labeyrie2006ltc}. The
CBS effect has also been studied on acoustic
\cite{Bayer_93,Tourin_97,de_Rosny_05} and seismic waves
\cite{Larose_2004}.

In most of the CBS experiments done  in the optical domain, the
angular distribution of the backscattered light is measured by
using an incoherent detection method. The signal that is detected
by a photomultiplier, a photodiode or a CCD camera is proportional
to the optical field intensity. In this letter we propose to study
the CBS effect more precisely by using a coherent detection method
\cite{pitter1997hde}. The backscattered field is summed with a
coherent reference field in order to detect the interference
pattern of the two fields. The detection is thus sensitive to the
phase of the backscattered field.

In this letter, we propose a detection scheme for the CBS effect,
which involves a heterodyne detection of the light on a 2D
detector (CCD camera). Since the pixel to pixel relative phase of
the detected backscattered field is accurately measured, we record
the digital hologram of the field \cite{LeClerc2000}. By adjusting
the frequency of the heterodyne detection local oscillator, we are
able to precisely select the frequency of the backscattered field
that is detected. We are thus able to measure the backscattered
light average intensity as a function of both the propagation
direction (as done commonly) and the frequency i.e. as a function
of both the wave vector direction $\textbf{k}/|\textbf{k}|$ and
the modulus $|\textbf{k}|$. The CBS effect can then be observed
with a backscattered field whose frequency is shifted with respect
to the incident incoming field. Since this shift is a Doppler
shift which is related to the motion of the scatterers, we can
study the influence of the motion of the scatterers on the CBS
signal. We will see in particular that the angular width (and more
generally the angular shape) of the enhanced backscattered cone
depends on the frequency offset.

\begin{figure}[]
\begin{center}
\includegraphics[width = 8.5 cm,keepaspectratio=true]{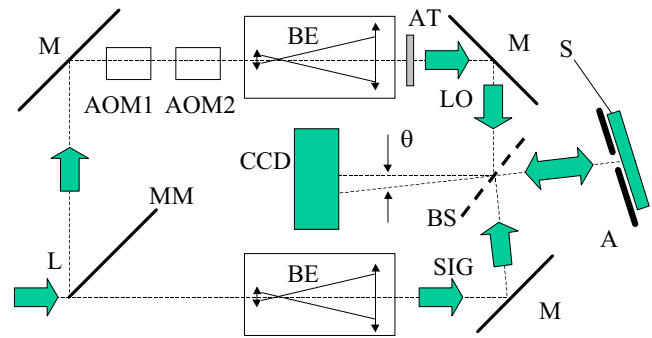}
\caption{Experimental Setup. L: main laser, MM: moving mirror, M:
mirror, LO and SIG: local oscillator and signal arms, BE: beam
expander, AT: attenuator, AOM1, AOM2: acousto optic modulators, ,
BS: beam splitter, S: cell filled with the diffusing suspension,
A: rectangular aperture, CCD: CCD camera. } \label{fig_setup}
\end{center}
\end{figure}

Fig \ref{fig_setup} shows the experimental  setup. The $780$ nm,
$50$ mW beam of the laser L (Sanyo DL-7140-201, current: $95$ mA)
is split in two beams LO and SIG, which are both vertically
linearly polarized. The main beam SIG ($>90 \%$ of the power,
frequency $\nu_L$) illuminates the diffusing sample S ($7$ cm high
$\times 8$ cm width $\times 1$ cm depth cell which is filled with
a pure or a diluted intralipid suspension: Endolipide $20\%$; B.
Braun Medical SA, diffusion anisotropy factor $g=0.65$ at
$\lambda=780 nm$ \cite{van_Staveren_91}). To detect the S
backscattered light signal, the LO beam ($<10 \%$ of the power) is
mixed to the signal beam with the beam splitter BS. The LO + SIG
interference pattern (or heterodyne beat) is recorded by a CCD
camera (PCO Pixelfly: $12$ bits, $\nu_{CCD}=12.5$ Hz exposure time
$20$ ms with $1280 \times 1024$ pixels of $6.7 \times 6.7 ~ \mu$m)
and transferred to a PC computer.

To avoid saturation of the CCD camera the LO beam is attenuated by
the attenuator AT. The power is finely adjusted by moving the
mirror MM. By using the AOM1 and AOM2 acousto optic modulators
(Crystal Technology: $\nu_{AOM1,2} \simeq 80 MHz$), the LO beam
frequency $\nu_{LO}$ can be freely adjusted: $\nu_{LO}=\nu_L +
(\nu_{AOM1}-\nu_{AOM2})$. Two numerical synthesizers with the same
$50$ MHz quartz reference clock provide the  $\nu_{AOM1,2}$
radiofrequency (RF) signals. Two beam expanders BE ($\times 25$:
Spectra Physics model 334) enlarge the LO and SIG beams. The BE
focus are adjusted to get a plane wave LO beam on the CCD, and a
plane SIG illumination beam on the sample.


To get the complex field $E$ at frequency $\nu$, we use the
4-phase heterodyne variant of the phase-shifting method
\cite{LeClerc2000,LeClerc_2001}. We record sequences of 4 CCD
images  $I_i$  at $t_i= t_0+ i / \nu_{CCD}$ with $i=0..3$. $E$
interferes with the local oscillator field $E_{LO}$, and
$I_i=|E_{LO}e^{j 2 \pi \nu_{LO} t_i} + E e^{j 2 \pi \nu t_i} |^2 $
where $j^2=-1$. We can choose $\nu_{LO}=\nu - \nu_{CCD}/4$, and
$|E_{LO}| \gg |E|$. $E$ is then given by the simple 4-phase
equation $E =a [(I_0-I_2)+ j(I_1-I_3)]$ where $a$ is a constant.
Due to the motion of the scatterers, the backscattered light
frequency spectrum is broad, and the 4-phase equation performs the
heterodyne detection on the two sidebands frequencies
$\nu^\pm=\nu_{LO}\pm \nu_{CCD}/4$, yielding to the field
components $E(\nu^\pm)$. The 4-phase equation should be thus
rewritten as: $ E(\nu^+) + E^*(\nu^-)=(I_0-I_2)+j.(I_1-I_3)$
\cite{Gross_03}. In the holographic reconstruction, the 2
components $E(\nu^+)$ and $E^*(\nu^-)$ yield to 2 different images
corresponding to the holographic grating orders $\pm 1$, which may
spatially overlap. To spatially separate the images, we have (i)
reduced the pupil size by using a black rectangular aperture A
($1$ cm width $\times 5$ cm high), (ii) moved the sample at a
respective distance $D_{sample}\simeq 25$ cm to the CCD, and (iii)
shifted the LO beam \emph{off axis} ($\theta$ on
Fig.\ref{fig_setup}) \cite{Schnars_Jueptner_94} .

%

\begin{figure}[]
\begin{center}
\includegraphics[width = 8.5 cm,keepaspectratio=true]{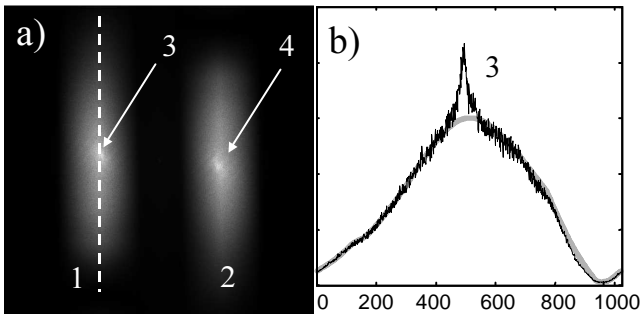}
\caption{A) Linear scale $1024\times 1024$ calculated  image  of the average k-space field
intensity $|\tilde E|^2$ for $\nu_{LO}=\nu_L$. One pixel is $1.13 \times 10^{-4}$ radian.  B) Cut
of the image along the dashed line (black curve). Angular gain of the experiment (solid grey line).
} \label{fig_image1}
\end{center}
\end{figure}

We have recorded  $N=256$ series of 4 images  ($4\times N$ images)
with the 1 cm thick  cell filled with Intralipid $10\%$ (optical
depth $\sim 20$ ${l_s}^*$  where ${l_s}^*$ is the transport mean
free path). The LO beam is unshifted in frequency (i.e.
$\nu_{LO}=\nu_L$), and the CCD acquisition time is $20$ ms per
image. For each series, we have calculated the real space field
$E(x,y)$ by the  4-phases equation, and the k-space field by 2D
Fast Fourier Transform (FFT): $\tilde E(k_x,k_y) = FFT ~E(x,y)$.
The angular distribution of the field intensity is then obtained
by summing $|\tilde E|^2$ over the $N$ series of images.
Fig.\ref{fig_image1}A) shows the k-space field calculated average
intensity $|\tilde E|^2$. Calculations are performed on a
$1024\times 1024$ matrices obtained from the $1280\times 1024$ CCD
raw data after truncation. From now on, we will consider the
angular size $\alpha_0$ of the k-space pixels as the units for the
angle. Due to the FFT we have: $\alpha_0 = \lambda/(1024 \times
d_{pixel}) =1.13 \times 10^{-4}$ radian.

The two vertical oblong  spots (1,2) seen on Fig.\ref{fig_image1}A
are the two grating orders ($\pm 1$) images of the sample.  Due to
the  off-axis configuration, images (1) and (2) do not overlap.
The images are blurred (the images are oblong while the aperture A
is rectangular) since the holographic reconstruction (FFT) is done
at infinite distance while the object is located at a finite
distance $D_{sample}$. The (1,2) spots also represent the angular
distribution of the backscattered light, and we can see narrow
brighter spots (3,4) in the middle of the spots (1,2) that are the
CBS enhanced cones. Due to the aperture A and moreover the fact
that the sample is quite far away from the CCD, the light, which
is backscattered far from the illumination axis, does not reach
the CCD. This means that the angular detection efficiency of our
setup is not flat (the wide spots 1 and 2 do not fill the
k-space). To analyze the shape of the CBS cone, we have made a cut
along the vertical line that passes in the center ($x_0,y_0$)  of
the CBS white spot (dashed line of Fig.\ref{fig_image1} A) and we
have plotted the CBS signal (black curve of Fig.\ref{fig_image1}
B). To improve the SNR (signal to noise ratio), the plotted curve
is obtained by summing over the lines $x=x_0-5$ to $x_0 + 5$. The
CBS peak (3) is clearly seen, but, as mentioned above, the
background is not flat.

We have corrected the signal from the background distortion  by
using the measured holographic data to calculate the complex field
image of the sample in the aperture A plane. From that calculation
we got the position of the aperture A   i.e. $D_{sample}$, the
shape of the illumination zone, and the local variations of the
illumination intensity. To get holographic signal $|\tilde E|^2$
that we should obtain if there is no CBS effect (i.e. if the
photons are scattered by the sample in all directions with equal
probability), we have calculated the blurred intensity image of
the objet that is expected with a focusing at infinite distance.
We have by this way the expected background signal i.e. the
efficiency of our setup for detecting the backscattered light in a
given angular direction. This quantity is plotted as a grey solid
line on Fig.\ref{fig_image1}B, which agrees  with the observed
shape of the CBS background.

\begin{figure}[htb]
\begin{center}
\includegraphics[width = 8.5 cm,keepaspectratio=true]{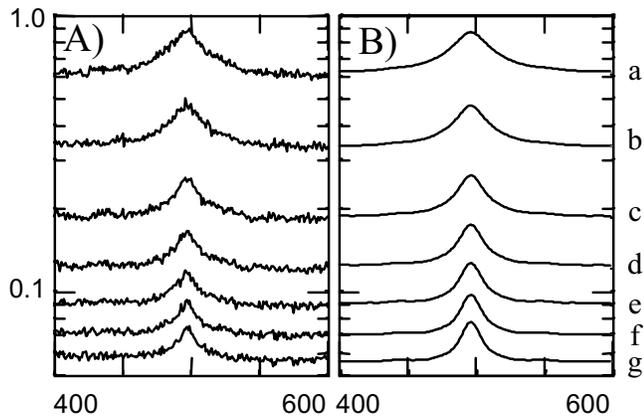}
\caption{ Angular shapes of the CBS peak for
$|\nu_{LO}-\nu_L|=0$(a), $100$(b), $200$(c), $300$(d), $400$(e),
$500$(f) and $600$ Hz(g). The curves are obtained by summing over
the $x=x_0-5$ to $x_0 + 5$ lines. To make all curves visible on
the same diagram, logarithmic scale is used on the $y$ vertical
axis ($y=0.05$ to $1.0$). Horizontal axis is angle ($1.13 \times
10^{-4}$ radian per point). A) Experimental results. B)
Theoretical predictions.}. \label{fig_curves}
\end{center}
\end{figure}

The heterodyne detection technique allows the study of new
physical effects since the frequency of the backscattered light
that is detected can be finely adjusted  within the brownian
spectrum. We have repeated the previous  experiment  for different
LO frequencies: $\nu_{LO}-\nu_L= + 0 $, $+ 100$ ...$ + 600$ Hz
(curve a, b, ...g of Fig.\ref{fig_curves}A).  To correct the
angular detection efficiency of our setup, the measured signal
(solid line of Fig.\ref{fig_image1}B) is divided by the expected
background signal (heavy grey line), which is calculated from the
$\nu_{LO}=\nu_L$ experimental data, in order to gain  the highest
SNR (signal to noise ratio) measurement. Fig. \ref{fig_curves}A
shows the CBS cone angular shape obtained by this way. The
detection efficiency correction is seen to be efficient and the
background is flat. As expected, the CBS cone signal and the
background both decrease when when $|\nu_{LO}-\nu_L|$ increases
(i.e. when the detection is done in the tail of the brownian
spectrum). Moreover, one observes a new physical effect, the
angular width of the CBS cone also decreases  with
$|\nu_{LO}-\nu_L|$ (curve a to curve g).

This effect can be understood quite simply.  By changing the
frequency of the detection  (i.e. $\nu_{LO}$), one changes the
distribution of travel paths that contribute to the signal. Since
each scattering event broadens the photon frequency spectrum, one
selects, by pushing $\nu_{LO}$ in the tail of the brownian
spectrum, travel paths with many scattering events (to get photons
far from the center of the brownian spectrum, one needs more
scattering events), i.e. longer travel paths. This corresponds to
a narrowing of the CBS peak.

\begin{figure}[]
\begin{center}
\includegraphics[width = 8.5 cm,keepaspectratio=true]{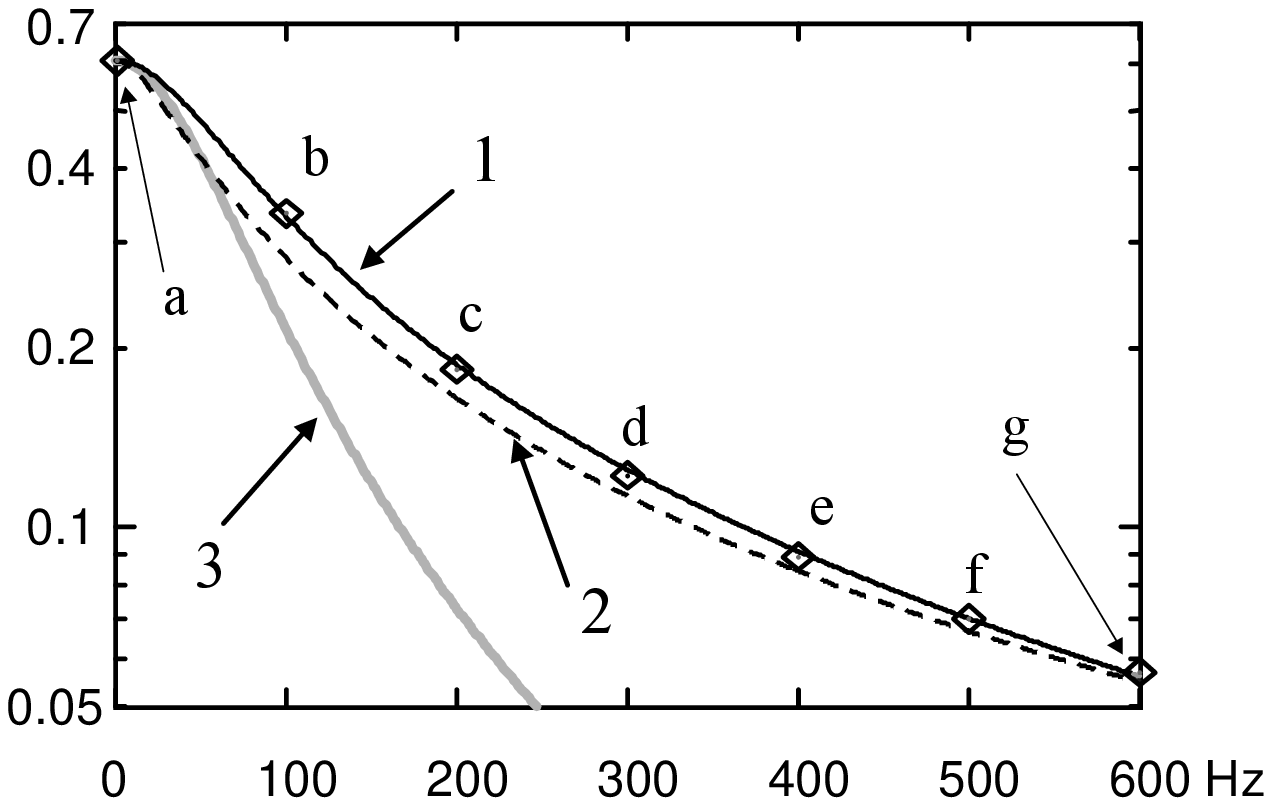}
\caption{Frequency spectrum of the backscattered light
(logarithmic scale vertical axis). Points are the CBS background
measured on curves a to g of Fig.\ref{fig_curves}A. Curve 1 (solid
line) is the brownian theoretical shape calculated by Monte Carlo,
curve 2 (dashed line) results from \cite{Pine_90} ), curve 3
(solid grey line) is the single scatterer Lorentzian. Horizontal
axis is $\nu_{LO}-\nu_L$.} \label{fig_brownian}
\end{center}
\end{figure}

To make a quantitative analysis of this effect, we have measured
the brownian diffusion constant $D_B$ of the intralipid particules
by considering that the CBS background measured on
Fig.\ref{fig_curves}A yields the shape of the brownian frequency
spectrum. In the case of the backscattering by single diffusor,
this spectrum is a Lorentzian of width $4k^2D_B$
\cite{berne_pecora_1976}. In the multi diffusion case,  each
travel path yields a Lorentzian of width  $D_B \Sigma_i
|\textbf{q}_i|^2$ (where $\Sigma_i$ is the sommation over the
scattering events $i$, and $\textbf{q}_i=
\textbf{k}_{i}-\textbf{k}_{i-1}$ is, for each event, the incident
versus exiting difference of wavevectors \cite{Pine_90}). The
spectrum is then a sum of Lorentzians. By considering a set of
10000 travel paths, we have calculated the shape of the multi
diffusion spectrum (Fig.\ref{fig_brownian} curve 1). The paths are
obtained by simple (scalar approximation, Henyey-Greenstein
scattering function) Monte Carlo simulation \cite{Jacques_98} with
$g=0.65$. The multi diffusion spectrum (curve 1) has been compared
with the single scatterer spectrum (curve 3), and  with the
diffusion wave spectroscopy predictions (curve 2). This one  is
the Fourier transform of $g_1(t)=e^{-\gamma(6 k^2 D_B t)^{1/2}}$
\cite{Pine_90} with $\gamma=1.9$ (i.e. for parallel polarisation
and $g=0.65$). We have adjusted the frequency scale to make curves
1 to fit the measured point (a to g). The width of curve 3
($4k^2D_B= 74$ Hz) yields  $D_B$.

\begin{figure}[]
\begin{center}
\includegraphics[width = 8 cm,keepaspectratio=true]{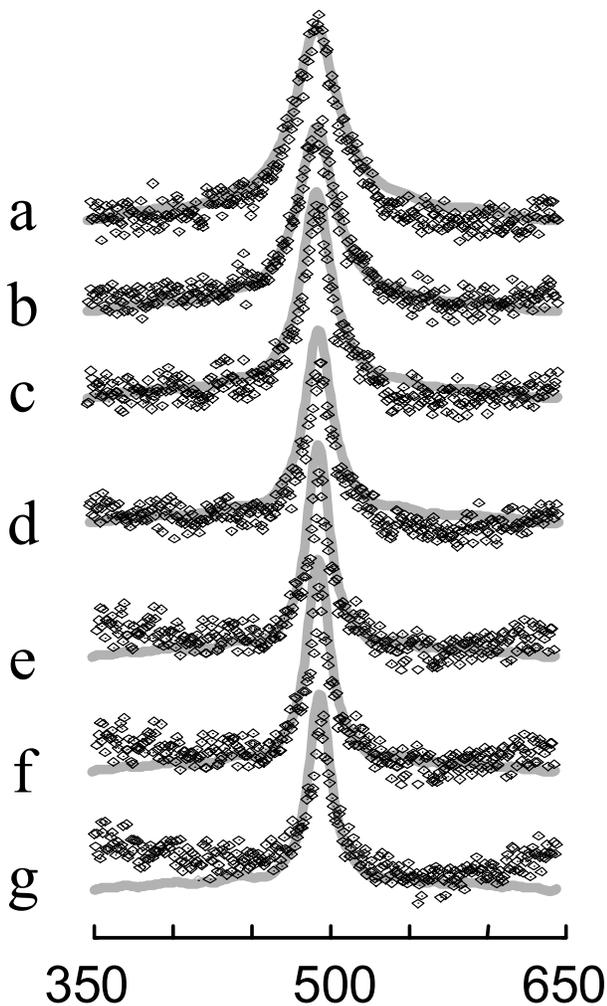}
\caption{Same as Fig.\ref{fig_curves} but  with arbitrary units linear scale vertical axis. The
points are experimental results, the solid grey curves the theoretical predictions.}
\label{fig_curves_detail}
\end{center}
\end{figure}
%
%
%

We have used our  Monte Carlo set of paths  to calculate the
angular shape of the CBS cone. For each  path, the photons may
travel forward  (from the first to the last scatterer) or in
reverse. The interference of the two paths  yield  a cosine
angular intensity distribution analog to a Young fringe pattern.
By summing these fringes over the paths, one gets the CBS enhanced
cone. This  intuitive idea dates to the CBS original papers
\cite{albada_85,Wolf_85}, and is illustrated by a web java applet
\footnote{http://cops.tnw.utwente.nl/education/5$\_$oc$\_$cohback.html}.
We have generalized this calculation by considering the angular
and frequency distribution of light for each path, and then
summing over the paths.
%
%
The CBS curves of Fig.\ref{fig_curves} correspond to the 2D cuts
(angle, frequency) distribution along the $\nu_{LO}-\nu_L=0$, 100,
...600 Hz lines. We have plotted on Fig.\ref{fig_curves}B the
curves that result from the random walk calculation. To make a
good theory versus experiment comparison, the
Fig.\ref{fig_curves}B curves are obtained by summing over the 11
cuts  made on the experimental data ($x=x_0-5$ to $x_0 + 5$). The
angular position of the CBS cone, the width of the CBS peak, and
the CBS enhanced factor $\eta$ have been determined from the curve
a of Fig\ref{fig_curves}A ($\nu_{LO}=\nu_L$). The other curves (b
to g) are then obtained without any free parameters.

As seen, the experimental results (Fig.\ref{fig_curves}A)  are in
good agreement with the theoretical predictions
(Fig.\ref{fig_curves}B). The CBS enhancement factor ($\eta=1.48$)
is lower than 2 because the linear polarizations are parallel, and
because we have summed over 11 parallel lines. With good optical
component and careful matching of the BE focus, our system could
be diffraction limited by the CCD area. In that case, the angular
resolution should exactly be one pixel of the k-space. We have
plotted the CBS curves with a linear scale for the vertical axis.
For each curve (a to g) the scale is arbitrarily adjusted.

In this letter, we report the first observation of a dependence of
the coherent backscattering (CBS) enhanced cone with the frequency
of the backscattered photons. The narrowing of the CBS cone that
is observed in the tail of the brownian spectrum is in good
agreement with a simple model which describes the effect of the
travel path distribution. We do not observe any break of
reciprocity because the scatterers do not move during the travel
of the photon and the scattering cross section does not depend on
the motion of the scatterers (as it could do in cold atom
experiments \cite{labeyrie2006ltc}). These results are made
possible by the development of a new detection method of the
backscattered photons, which is based on heterodyne detection of
the photons with a CCD camera. This method, which allows the study
of both the angular and spectral properties of the backscattered
light, opens the way to the observation of new physical effects in
disordered systems. The authors thank D.Delande, A. Tourin,  J. de
Rosny, G. Montambaux and D. Bonn for fruitful discussions. Authors
acknowledge also french ANR for its support.


\end{document}